\documentclass[prb,amsmath,amssymb]{revtex4}

\usepackage[dvips]{graphicx}
\usepackage{bm}

\begin{document}

\title{Colossal spin fluctuations in a molecular quantum dot magnet\\ with ferromagnetic electrodes}

\author{Thibaut Jonckheere$^{1}$, Ken-Ichiro Imura$^2$, and Thierry Martin$^{1,3}$}
\affiliation{
${}^1$~Centre de Physique Th\'eorique, Case 907 Luminy, F-13288 Marseille Cedex 9, France\\
${}^2$~Institute for Solid State Physics, University of Tokyo, Kashiwanoha 5-1-5, Kashiwa 277-, Japan\\
${}^3$~Universit\'e de la M\'edit\'erann\'ee, F-13288 Marseille Cedex 9, France
}

\begin{abstract}
We study electronic transport through a magnetic molecule with an intrinsic spin $S$ coupled
to two magnetic electrodes, in the incoherent regime. The molecule is modeled as a single resonant level
with large Coulomb repulsion (no double occupancy). The molecular spin
is isotropic and it interacts with the electronic spin through an exchange interaction.
Using an alternative method to the usual master equation approach, we are able to obtain 
analytical formulas for various physical quantities of interest, 
such as the mean current and the current fluctuations, 
but also the mean value of $J_z$ -the $z$ component of the total spin on the molecule- and its fluctuations. 
This allows us to understand 
how the electronic current between the magnetized electrodes
can control the polarization of the molecular spin. We observe in particular that the fluctuations
of $J_z$ reach unexpectedly high values.
\end{abstract}

\maketitle

\section{Introduction} 

\begin{figure}
\centerline{\includegraphics[width=8.cm]{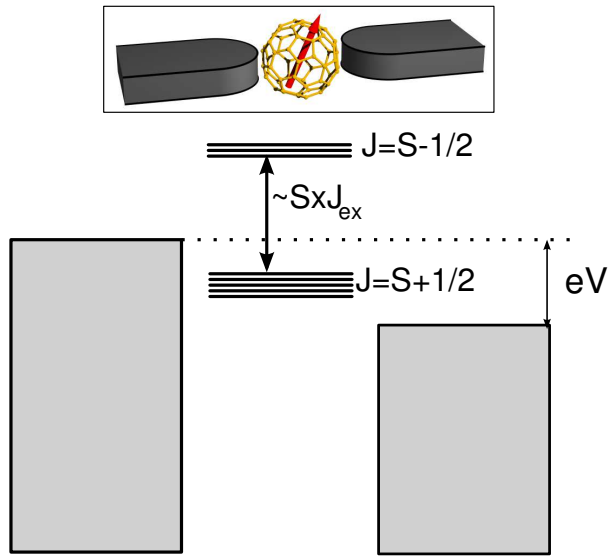}}
\caption{Energy diagram of the system: because of the exchange interaction between the molecular spin $S$ and the electron spin,
the occupied dot level is split between the $J=S-\frac{1}{2}$ levels (with $J_z$ ranging from
$-(S-\frac{1}{2})$ to $(S-\frac{1}{2})$) and the $J=S+\frac{1}{2}$ levels (with $J_z$ ranging from $-S-\frac{1}{2}$ to $S+\frac{1}{2}$), 
with a spliting $\sim J_{ex} S$. We chose
the chemical potential of the electrodes such that the $J=S+\frac{1}{2}$ levels only are in the bias window. The inset shows a schematic view of a possible 
experimental realization: a magnetic atom with spin $S$ trapped inside a $C_{60}$ molecule, which is placed between two electrodes. }
\label{fig:levels}
\end{figure}

Molecular spintronics is at the convergence of two recent and rapidly developping fields. On the one hand,
molecular electronics, \cite{molecular1}
where individual molecules are connected  to electrodes of different nature, and
the effect of the various molecular degrees of freedom on the electronic transport can be studied and possibly engineered. 
\cite{molecular2}
On the other hand, 
spintronics, where the focus is placed on the electronic spin as a new useful degree of freedom.
Magnetic molecules - molecules having an intrinsic spin, possibly large \cite{Mn_12} - 
play, of course, an important role in molecular spintronics. 
Several transport experiments have been performed in the past years on such magnetic molecules - 
specially on molecular magnets like $Mn_{12}$ derivatives, which are molecules with a large spin anisotropy, tending to align the spin along an easy-axis.
\cite{heersche}
Theoretical calculations on transport in the incoherent regime for these molecules
have been done, specially in the case where the electrodes have magnetic properties.
\cite{elste_timm2006}
Other magnetic molecules are spin isotropic,
and some work has been done already to study electronic transport with such
spin-isotropic magnetic molecule.
\cite{elste_timm2005}
In Ref. [\onlinecite{fcs2007}], the full counting statistics (FCS) for such a molecule 
placed between non-magnetic electrodes has been obtained.
However, to the best of our knowledge, no work has been devoted to the electronic transport
between ferromagnetic electrodes through a spin-isotropic magnetic molecule;
one of the aims of this paper to study this problem. Experimentaly, transport through
such a spin-isotropic magnetic molecule can be obtained for example with a magnetic atom
trapped inside a $C_{60}$ molecule, which is placed between two electrodes (see Fig.~\ref{fig:levels}).\cite{kasumov2005}

This paper focuses on the transport through a spin-isotropic magnetic molecule, 
in the regime of weak coupling to the leads.
The primary goal is to compute the current, its zero-frequency fluctuations 
and more importantly to analyze the fluctuations of the total spin on the molecule.
This is an important issue because of the mutual influence between the electronic current passing through the molecule and
the molecular spin. We consider ferromagnetic electrodes with collinear alignments of electrodes (parallel
or anti-parallel, or situation with only one polarized electrode).\cite{noncollinear} The system
displays a rich variety of behaviors:  on the one hand the suppression of the current by spin-blockade
and, on the other hand, unusually large fluctuations of the molecular spin.

We are considering the temperature regime in which successive 
tunneling events through the dot are all incoherent (incoherent tunneling regime),
and describe them as a Markovian process.
Such a situation is realized at a temperature $\Theta$ which is much higher than the typical 
energy scale determined by $\Gamma$, i.e., $\hbar\Gamma\ll k_B \Theta$.
We model the molecule as a quantum dot with a single resonant level, with infinite 
Coulomb repulsion (no double occupancy).
The molecular spin $\vec{S}$ , and the electronic spin
on the dot level $\vec{\sigma}$, interact through an exchange interaction,
$-J_{ex} \vec{S} \cdot \vec{\sigma}$.
The dot level is, therefore, split into two levels, corresponding to two eigenvalues
of the total angular momentum, $J=S\pm 1/2$. 
For simplicity, we will work at temperatures much smaller than 
the applied bias voltage, at which electron transport happens {\it only in one direction},
and the bias window is infinitely sharp. 
\cite{T-dep}
We also choose the chemical potentials of the electrodes such that 
only the $J=S+1/2$ spin sector is in the bias window,
and only states in this energy level take part in the transport (see Fig.~\ref{fig:levels}).

The basic mechanism at work here is the exchange of spin between the itinerant 
electrons and the molecule: an incident spin-up electron, for example, 
can be collected as a spin-down electron (spin-flip), 
if the molecular spin has its polarization along the reference axis increased by one. 
Magnetic electrodes, with different densities of states for spin up/down electrons, 
leading to spin-depedent tunneling rates:
$\Gamma_{L,R}^\uparrow=\Gamma_{L,R}(1+p_{L,R})/2$, 
$\Gamma_{L,R}^\downarrow=\Gamma_{L,R}(1-p_{L,R})/2$, 
can thus induce polarization of the molecular spin ($p_{L,R} \in [-1,1]$ is the polarization of the $L/R$ electrode,
and $\Gamma_{L,R}$ is the tunneling rate to the $L/R$ electrode).

The standard method in the incoherent tunneling regime is to use master equations.
This approach has been successfully implemented to compute numerically the current and the noise 
through such molecular systems.\cite{elste_timm2005,elste_timm2006} Recently, analytical results were obtained
for the full counting statistics in the case of non-magnetic electrodes.\cite{fcs2007}
We use here an alternative method which allow us to obtain {\it analytical formulas} 
for the case of magnetic electrodes, for the current $I$, the charge $Q$, the total spin $J_z$
and the fluctuations of these quantities.  
This method has been introduced by Korotkov, for computing numerically fluctuations 
 in the single-electron transistor.
\cite{korotkov1994}
It uses a Langevin approach, where the transport process is seen as random 
sequential jumps between neighbouring system states. 
We have extended this method, in order to obtain analytical results for the present problem. 

The paper is organized as follows. 
In section II we present a concise but self-contained explanation of the method. 
Section III presents and discusses the results we have obtained in the case of 
a molecular spin $S=1/2$, for the mean current and its fluctuations,
and for the $z$ component of the total spin $J_z$ and its fluctuations. 
Section IV discusses how the results are modified
in the case of higher spins. Another method of calculation which can be used to obtain 
the same analytical results is shortly explained in section V, and section VI gives the conclusion. 
A few appendices contain some lengthy formulas, and analytical results for
molecular spin $S=1$.

\section{The segment picture}
We give a short self-contained derivation of the method, only stressing the points which are different from 
the original work.\cite{korotkov1994}

\subsection{General formulation}
The time evolution is divided in terms of segments: 
a segment $\zeta$ is defined as a series of random processes which begins with
a reference state and finishes with the same state. This reference state is arbitrary, and all the physical
quantities are of course independent of the choice of this state.
As the time evolution
is given by a Markovian series of random transitions, two different segments are totally independent, 
and any time integral used to compute average or fluctuations can be written in terms
of average over the segments. 
In our model, a state $|\alpha\rangle$ can be characterized by 
the occupation number of the dot level $Q$, the total angular momentum $J$ and its $z$-component$J_z$,
i.e.,
$|\alpha\rangle=|Q,J,J_z\rangle$.
A segment of length $M$, starting and finishing with state $\alpha_0$ is thus  defined by the sequence
$\alpha_0\rightarrow\alpha_1\rightarrow\alpha_2\rightarrow\cdots
\rightarrow\alpha_{M-1}\rightarrow\alpha_0$, and by the duration of each step.
The total duration of the segment is
$\tau[\zeta]=\sum_{m=0}^{M-1}\tau_m$, where $\tau_m$ is the time the system stays in the state $\alpha_m$.

Considering a random variable $X(t)$, we will compute its average $\bar{X}$ over the measurement time, i.e.,
\begin{equation}
\bar{X}\equiv {1\over T}\left\langle
\int_0^T dt X(t)\right\rangle,
\label{Xbar}
\end{equation}
and its fluctuations $S_{XX}$:
\begin{equation}
S_{XX}\equiv {2\over T}\left\langle
\left(\int_0^T dt \left(X(t)-\bar{X}\right)\right)^2
\right\rangle.
\label{cmom}
\end{equation}
Here $\langle \cdots\rangle$ represents a statistical average over
random Markovian process, and we take the measurement time $T$ sufficiently larger than all other time scales
of the system. 
For each physical quantity $X$, we define a function ${\cal X}[\zeta]$
which gives the time integral of this quantity over a given segment $\zeta$, e.g., for $J_z$, we define,
\begin{equation}
{\cal J}_z[\zeta]=\sum_{m=0}^{M-1}J_{z,m}\tau_m,
\label{jzxi}
\end{equation}
where $J_{z,m}$ is the value of $J_z$ in the state $\alpha_m$. 
We have similar expressions for all the other physical quantities which have a fixed value in a given state 
$\alpha_m$.
For the current operator $I$, as its time integration gives the transfered charge, we need to define
the function $k[\zeta]$ which is the number of electrons tranfered from the left to the right electrode during segment 
$\zeta$. 
With these functions, the time integral of $X(t)$ in Eq. (\ref{Xbar}) over the measurement time $T$
can be decomposed into contributions from $N$ successive segments,
$\{\zeta_1,\zeta_2,\cdots,\zeta_N\}$:
\begin{equation}
\int_0^T dt X(t)\rightarrow
\sum_{n=1}^N {\cal X}[\zeta_n].
\end{equation}
Taking it also into account that the different segments are independent, one can thus
replace the statistical average  with 
{\it an average over the segments};
\begin{equation}
\bar{X} = \frac{1}{T} 
\sum_{n=1}^N \left\langle{\cal X}[\zeta_n] \right\rangle 
= \frac{\langle {\cal X} \rangle}{\langle \tau \rangle},
\label{Xav}
\end{equation}
where $\langle \tau \rangle=T/N$ is the mean duration of a segment.
A segment $\zeta$ occurs with a probability $P[\zeta]$, giving
$\langle{\cal X}\rangle=\sum_\zeta {\cal X}[\zeta]P[\zeta]$.

For the fluctuations of $X$ one first notices that
\begin{equation}
\left(
\int_0^T dt \left(X(t)-\bar{X}\right)
\right)^2=
\sum_{n=1}^{N}\sum_{n'=1}^{N} \left({\cal X}[\zeta_n]-\bar{X} \tau[\zeta_n]\right)
                             \left ({\cal X}[\zeta_{n'}]-\bar{X} \tau[\zeta_{n'}] \right)
\label{decom}
\end{equation}
We note that $\zeta_n$ and $\zeta_{n'}$ are different segments of {\it the same
measurement}, i.e., the set $\{\zeta\}=\{\zeta_1,\zeta_2,\cdots,\zeta_N\}$ is common for
the sum over $n$ and $n'$, and the average is an average over the different segments sets $\{\zeta\}$.
As different segments are independent, the terms with $n\neq n'$ can be written as a product of two averages,
which clearly vanish, i.e., $\langle {\cal X}[\zeta_n]-\bar{X}\tau[\zeta_n]\rangle=0$.
As a result, one finds,
\begin{eqnarray}
S_{XX}&=&{2\over T}\left\langle
\left(\int_0^T dt \left(X(t)-\bar{X}\right)\right)^2
\right\rangle
=
{2\over T}
\sum_{n=1}^{N}
\left\langle
\left(
{\cal X}[\zeta_n]-
{\langle{\cal X}\rangle
\over
\langle\tau\rangle}
\tau[\zeta_n]
\right)^2
\right\rangle
\nonumber \\
&=&
{2\over\langle\tau\rangle}
\left[
\langle{\cal X}^2\rangle+
\langle\tau^2\rangle
\left(
{\langle{\cal X}\rangle
\over
\langle\tau\rangle}
\right)^2-
2\langle{\cal X}\tau\rangle
{\langle{\cal X}\rangle
\over
\langle\tau\rangle}
\right].
\label{Xnoise}
\end{eqnarray}
We can thus express the fluctuations $S_{XX}$ in terms of the averages
over the segments, such as,
$\langle\tau\rangle$, $\langle{\cal X}\rangle$, $\langle{\cal X}\tau\rangle$, etc. 
We note that these are averages over either linear (such as, $\langle\tau\rangle$, $\langle {\cal J}_z \rangle $),
or quadratic (such as, $\langle {\cal J}_z^2 \rangle$, $\langle {\cal J}_z \tau \rangle$)
functions. Note that centered moments of higher order (for example the third centered moment
 $\sim 1/T \langle (\int_0^T dt (X(t)-\bar{X}) )^3 \rangle$) cannot be easily obtained with this segment technique. Indeed,
the equivalent of Eq.~(\ref{Xnoise}) for a higher moment has terms involving averages over the last, incomplete, segment
of the time interval $[0,T]$. The contribution from this incomplete segment is negligible ($\sim 1/T$) for the first and second moment, but
is important for higher moments. As this last segment is incomplete, its statistics is different from the statistics of standard segments and
cannot be easily computed.

As is emphasized in Ref. [\onlinecite{korotkov1994}], the average over the segments can be done in two steps:
\begin{equation}
\langle{\cal X}\rangle=\langle{\cal X}\rangle_{1,2}=
\sum_\zeta \langle{\cal X}[\zeta]\rangle_1 P_2[\zeta].
\label{av12}
\end{equation}
The first average is carried out over the durations $\tau_0$,...,$\tau_{M-1}$ of an arbitrary sequence of states 
$\alpha_0\rightarrow\alpha_1\rightarrow\cdots\rightarrow\alpha_{M-1}\rightarrow\alpha_0$. 
This first average, which we denoted in Eq. (\ref{av12}) as $\langle\ldots\rangle_1 $,
is easy to perform, as $\tau_m$ is given by a Poissonian process, 
with a rate $\Gamma_m$, and we leave further details to Appendix A. 
The second step is an average over all the possible sequences of states,  with
the correct probability $P_2[\zeta]$ for each sequence. As we will show below, it is possible in our case to 
describe the whole set of sequences, and this second average can also be performed analytically.

\subsection{Construction of all possible segments --- case of molecular quantum dot magnet}
In order to perform the second average, we must identify the whole set of possible sequences for
the magnetic molecule system.
As we work in the limit of strong Coulomb blockade, the dot level can be occupied
at most by 1 electron, so we have $Q=0$ or $Q=1$.
For the empty dot, the total spin is simply given by the molecular spin, and specifying
the $z$ component of the spin determines the state completely ($S_z \in [-S,S]$),
 so an empty dot state  is given by $|0, S_z\rangle$. For the occupied dot ($Q=1$), the total spin is obtained by
the addition of the molecular spin and the spin of the electron occupying the dot, which gives $J=S \pm 1/2$. These two values
of the total spin correspond to two levels of the system, separated by an energy of order $J_{ex}$ (the value
of the exchange coupling between the spins).
 As explained before, we decide here to work in the ferromagnetic case, where 
the lower level is the one with $J=S +1/2$, and with the chemical potentials of the electrodes placed so that only 
this lower level is in the bias window; 
then the level $J=S-1/2$ plays no role in transport (it cannot be populated) and can be forgotten.
The occupied dot has thus a total spin $J=S+1/2$, and specifying the $z$ component again determines the state completely ($J_z \in
[-(S+1/2),S+1/2]$), so an occupied dot state is given by $|1, J_z\rangle$. 

For the reference state (which is the initial and final state of each segment), we choose an empty dot 
with spin maximally polarized along the $z$ axis: $|0, S_z=+S\rangle$.  
From this state, there are two basic sequences where a single electron is tranfered from
the left to the right electrode:
\begin{itemize}
\item(A) $|0,S_z=+S \rangle \to |1,J_z = S+1/2 \rangle \to |0, S_z=+S\rangle$
\item(B) $|0,S_z=+S \rangle \to |1,J_z = S-1/2 \rangle \to |0, S_z=+S\rangle$.
\end{itemize}

The two sequences (A) and (B) are the two simplest ones.
Clearly, the sequence A cannot be extended further, as the spin of the intermediate state is maximal. 
On the contrary, the sequence (B), which we will call {\it a basic sequence B}, can be extended 
by adding a subsequence starting and finishing at the intermediate state $|1,J_z=S-1/2\rangle$, 
and going only to lower values of $S_z$. One can add 
\begin{eqnarray}
&& \quad \bullet \;i_1 \mbox{ times the subsequence } v_1 \nonumber \\
&&   |1,J_z=S-1/2\rangle \to |0,S_z=S-1\rangle \to |1, J_z=S-1/2\rangle \nonumber \\
&& \quad \bullet \;i_2 \mbox{ times the subsequence } v_2 \nonumber \\
&&  |0,S_z=S-1\rangle \to |1,J_z=S-3/2\rangle \to |0, S_z=S-1\rangle \nonumber \\
&& \quad \quad \quad \quad \cdots  \nonumber \\
&& \quad \bullet \;i_{4S} \mbox{ times the subsequence } v_{4S} \nonumber\\
&& |0, S_z=-S\rangle \to |1,J_z=-S-1/2\rangle \to |0,S_z=-S\rangle \nonumber
\label{eq:subsequences}
\end{eqnarray}   
\begin{figure*}[!]
\begin{minipage}{12cm}
\includegraphics[width=12cm]{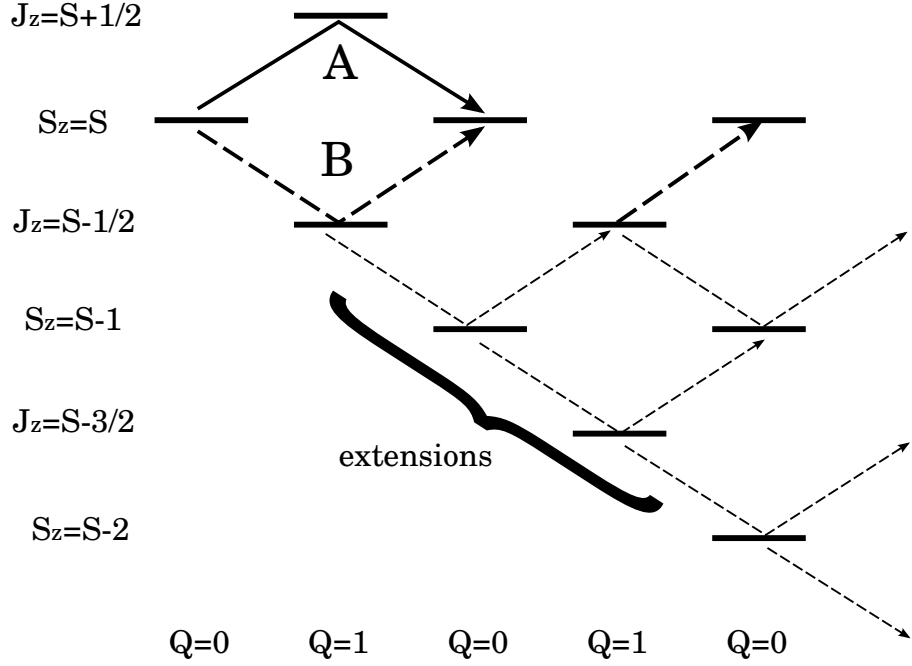}
\end{minipage}
\caption{Construction of all the possible sequences - type A and type B$+$extensions.
A sequence starts and ends in the reference state, $Q=0,S_z=S$. Sequence A (solid line) correspond to the tunneling
of a spin-up electron from the left to the right lead. Sequence B (dashed line) to the tunneling of
a spin-down electron from left to right lead. Sequence B can be extended
by attaching subsequences starting and ending from the state  $Q=1,J_z=S-1/2$ (thin dashed lines), forming
longer sequences where several electrons are transmitted, 
and where the molecular spin goes through intermediate states with $S_z<S$ and $J_z<S-1/2$.
}
\label{fig:segments}
\end{figure*}
Combining the $4S$ subsequences to the basic segment B, 
with arbitrary repetition of each subsequence, 
one can {\it construct} all the possible sequences of type B (see also Fig.~\ref{fig:segments}). 
We use the notation $B^*(i_1,i_2,\dots,i_{4S})$ to represent the type-B sequence composed
of the basic sequence B combined with $i_1$ times the subsequence $v_1$, $i_2$ times the subsequence $v_2$,
etc. The basic sequence B is then simply $B^*(0,0,\dots,0)$. 
The whole set of type-B sequences, plus the simple sequence A, give all the possible sequences.

The probability with which a given sequence occurs is given by the product of the probabilities of the transitions 
forming that sequence. 
The probability of a transition is given by the transition rate divided by the total transition rate of
the initial state. One must here distinguish between transitions starting from an empty dot ($Q=0$) and
transition starting from an occupied dot ($Q=1$).
An empty dot ($Q=0$) with a molecular spin $S_z$ is subject to
two types of transitions: tunnel-in of an electron from the
left electrode with either up or down spin, which brings the dot to the state $J_z=S_z+1/2$ or
$J_z=S_z-1/2$. The transition rates $\Gamma_{Q=0}^+(S_z)$ and $\Gamma_{Q=0}^-(S_z)$ 
for the above two processes can be calculated, using Fermi's
golden rule and Clebsh-Gordan coefficients:
\begin{equation}
\Gamma_{Q=0}^{\pm}(S_z)=\Gamma_L \frac{1\pm p_L}{2}  \frac{S+1\pm S_z}{2S+1}.
\label{q0def1}
\end{equation}
Using the tunneling rates $\Gamma_{Q=0}^\pm(S_z)$, one can express
the probability $P_{Q=0}^\pm(S_z)$
with which the system jumps onto either of the two final states:
\begin{equation}
P_{Q=0}^\pm(S_z)={\Gamma_{Q=0}^\pm(S_z)\over
\Gamma_{Q=0}^+(S_z)+\Gamma_{Q=0}^-(S_z)}.
\label{q0def2}
\end{equation}
For an occupied dot ($Q=1$), with the molecular spin state $J_z$, tunnel-out of an electron
with either spin-down or spin-up brings the dot to the state $S_z=J_z+1/2$ or $S_z=J_z-1/2$, with
the rates
\begin{equation}
\Gamma_{Q=1}^{\pm}(J_z) = \Gamma_R \frac{1\mp p_R}{2} \frac{S+1/2\mp J_z}{2S+1} ,
\end{equation}
giving the probability
\begin{equation}
P_{Q=1}^\pm(J_z) = {\Gamma_{Q=1}^\pm(J_z)\over
\Gamma_{Q=1}^+(J_z)+\Gamma_{Q=1}^-(J_z)}.
\label{q1def}
\end{equation}
Using Eq. (\ref{q0def2},\ref{q1def}), one can express the probabilities of 
sequences A and B:
\begin{eqnarray}
P_2[A]&=&P_{Q=0}^+(S_z=S)\times P_{Q=1}^-(J_z=S+1/2),
\nonumber \\
P_2[B]&=&P_{Q=0}^-(S_z=S)\times P_{Q=1}^+(J_z=S-1/2).
\label{probA}
\end{eqnarray}
The subscript 2 is to recall that these probabilities are associated with
the second average (average over different sequences).
Similarly, extensions starting from an occupied dot or an empty dot occur with the probability,
\begin{eqnarray}
P_{\rm o}(J_z)&=&P_{Q=1}^-(J_z)\times P_{Q=0}^+(S_z=J_z-1/2) \nonumber \\
P_{\rm e}(S_z)&=&P_{Q=0}^-(S_z)\times P_{Q=1}^+(J_z=S_z-1/2)
\end{eqnarray}
where the subscript $o$ and $e$ stand for occupied and empty.
The probability of the sequence $B^*(i_1,i_2,\dots,i_{4S})$ is then
\begin{equation}
P_2[B^*(i_1,i_2,\dots,i_{4S})] = 
C^{i_2}_{i_1+i_2-1} \dots  C^{i_{4S}}_{i_{4S-1}+i_{4S}-1} \;
p_1^{i_1} p_2^{i_2}\cdots p_{4S}^{i_{4S}} \;
P_2[B],
\label{probB}
\end{equation}
where $p_{2l-1}\equiv P_{\rm o}(J_z=S-(2l-1)/2)$,
$p_{2l} \equiv P_{\rm e}(S_z=S-l)$.
In this expression, the combinatorial factors ($C^j_i = i!/(j! (i-j)!$) count the number of different sequences corresponding to the
set $(i_1,i_2,\dots,i_{4S})$ because of the possible permutations of the subsequences. 
There are $C^{i_n}_{i_{n-1} +i_{n} -1}$ different
possibilities to ``attach'' the $i_l$ subsequences at level $l$ to 
one of the intermediate states of the $i_{l-1}$ subsequences at level $l-1$.
The second part of the expression is simply the product of the probabilities of all the subsequences.

Together with some explicit formulas for the first average (see Appendix A),
 Eqs. (\ref{probA},\ref{probB}) allow us to evaluate the averages over different segments $\zeta$ 
appearing in Eq. (\ref{av12}).
Evaluating all such averages, one finally finds the explicit formulas for 
$\bar{X}$ and $S_{XX}$, formally written as Eqs. (\ref{Xav}) and (\ref{Xnoise}).
Analytic results  for such quantities are listed in Tables I and II,
 as a function the polarization $p$ of the electrodes.
Some of such examples are also shown in Appendix B, along with some intermediate steps
in a specific case of $S=1/2$. Note that we are able to obtain such analytical formulas
 owing to identities involving summation on the binomial factors, like:
\begin{eqnarray}
\sum_{j=0}^{\infty} C^j_{i+j-1} \, x^j &=& \frac{x}{(1-x)^{(i+1)}} \nonumber \\
\sum_{j=0}^{\infty} C^j_{i+j-1} \, j \, x^j &=& \frac{x (1+i x)}{(1-x)^{(i+2)}}
\end{eqnarray}
In practice, the calculations are quite lengthy, but results are easily obtained using a symbolic computation software. 
In Appendix B, it is also shown that our analytic results are consistent with the cumulant 
generating function, obtained in Ref. [\onlinecite{fcs2007}], in the limit of non-magnetic electrodes: 
$p\rightarrow 0$.

\section{Results}
In this section, we present the results we have obtained for the various quantities.
We will focus on the mean current $\bar{I}$ and the current noise $S_{II}$,
on the mean charge on the dot $\bar{Q}$ and its fluctuations $S_{QQ}$,
and on the mean value of z-component of the spin on the dot, $\bar{J_z}$ and its fluctuations $S_{J_zJ_z}$. 
In order to see the effect of the leads magnetization,
we will show all these quantities as a function of the polarization 
of the leads. For simplicity, we choose 
to have a single parameter for the leads polarizations, 
and we have chosen four representative cases. In the first two cases, the two electrodes
are magnetic. The absolute value and the direction of the polarizations are the same in the two electrodes, but 
are either parallel ($P_L = P_R = p$, case noted P), or anti-parallel ($P_L = -P_R = p$, case AP).
In the two remaining cases, only one of the electrode is magnetic (with a polarization $p$),
 while the other one has no magnetic property. 
The polarized electrode can be either the left one, which is the source electrode (case LP), or the right one,
which is the drain electrode (case RP).

Table~\ref{table:ana12} gives the anaylitic formulas we have obtained, in the four different
cases for the leads polarization, for a molecular spin $S=1/2$ (Appendix~\ref{app:tableana1} shows a similar table
for the case $S=1$; we do not show any formula for a higher spin $S$ 
as they become too lengthy).
  For simplicity, we have chosen equal bare transition rates for the left and right electrodes, $\Gamma^{(0)}_L=\Gamma^{(0)}_R= 1$
(formulas with general transition rates can be obtained easily with the same method).
 As can be seen on the table, each analytic formula is given by a fraction of two polynomial in $p$, and containing only of even powers
 of $p$ (except for $\bar{J_z}$ where there is an additional factor $p$). 
 The orders of these polynomial, and their coefficient, increase
when the molecular spin is increased (see  Appendix~\ref{app:tableana1})

\begin{table*}[!]
\[
\begin{array}{|c||c|c|}
\hline
P &  \bar{I}  & \displaystyle{ 3/10 }   \\ \hline
AP &  \bar{I} & \displaystyle{ \frac{-5 p^4+2 p^2+3}{2 p^4+20 p^2+10} }   \\\hline
LP &  \bar{I} & \displaystyle{ \frac{p^2 + 3}{2 p^2 +10}}\\\hline
RP &  \bar{I} & \displaystyle{\frac{3(1-p^2)}{2(5-p^2)} }  \\\hline\hline 
P &  S_{II} & \displaystyle{\frac{125 p^2+39}{125-125 p^2}  } \\\hline
AP &  S_{II} & \displaystyle{\frac{43 p^{12}-318 p^{10}-463 p^8+12 p^6+413 p^4+274
   p^2+39}{\left(p^4+10 p^2+5\right)^3} }  \\\hline
LP &  S_{II} & \displaystyle{\frac{p^6+p^4+39 p^2+39}{\left(p^2+5\right)^3} } \\\hline
RP &  S_{II} & \displaystyle{\frac{3 \left(15 p^6-49 p^4+21
   p^2+13\right)}{\left(5-p^2\right)^3} }\\\hline\hline 
P & \bar{J_z} & \displaystyle{ 0 } \\ \hline
AP & \bar{J_z} & \displaystyle{ \frac{2 p \left(3 p^2+5\right)}{p^4+10 p^2+5} } \\ \hline
LP & \bar{J_z} & \displaystyle{ \frac{5 p}{p^2+5} } \\ \hline
RP & \bar{J_z} & \displaystyle{ -p } \\ \hline \hline
P & S_{J_zJ_z} & \displaystyle{ \frac{88}{5-5 p^2} } \\ \hline
AP & S_{J_zJ_z} & \displaystyle{\frac{8 (1-p^2) \left(-25 p^8+48 p^6+30 p^4-200
   p^2+275\right)}{\left(p^4+10 p^2+5\right)^3}} \\ \hline
LP & S_{J_zJ_z} & \displaystyle{\frac{8 \left(-2 p^6+71 p^4-340
   p^2+275\right)}{\left(p^2+5\right)^3}} \\ \hline
RP & S_{J_zJ_z} & \displaystyle{\frac{8 \left(8 p^4-19 p^2+11\right)}{5-p^2}} \\ \hline \hline 
P & \bar{Q} & \displaystyle{3/5 } \\ \hline
AP & \bar{Q} & \displaystyle{\frac{3 p^4+10 p^2+3}{p^4+10 p^2+5} } \\ \hline
LP & \bar{Q} & \displaystyle{\frac{p^2+3}{p^2+5} } \\ \hline
RP & \bar{Q} & \displaystyle{\frac{p^2+3}{5-p^2}} \\ \hline \hline
P & S_{QQ} &  \displaystyle{ \frac{96}{125 \left(1-p^2\right)} } \\ \hline
AP & S_{QQ} &  \displaystyle{\frac{32 (1-p^2) \left(15 p^8+40 p^6+38 p^4+32
   p^2+3\right)}{\left(p^4+10 p^2+5\right)^3} } \\ \hline
LP & S_{QQ} &  \displaystyle{\frac{32 \left(-p^4+2 p^2+3\right)}{\left(p^2+5\right)^3} } \\ \hline
RP & S_{QQ} &  \displaystyle{\frac{32 (1-p^2)\left(-3 p^4+8
   p^2+3\right)}{\left(5-p^2\right)^3}} \\ \hline
\end{array}
\]
\caption{The analytic formulas for the case of a molecular spin $S=1/2$, when only the
triplet state of the occupied dot (with a spin $S=1$) lies in the bias window. The first column shows the type of
polarizations in the leads : P for parallel ($P_L=P_R = p$), AP for anti-parallel ($P_L=-P_R=p$),
 LP ($P_L=p$, $P_R=0$) for left lead polarized only and RP ($P_R=p$, $P_L=0$) for right lead polarized only. 
The second column shows the quantity whose analytical formula is given in the third column}
\label{table:ana12}
\end{table*}

\subsection{Mean current and current fluctuations}

\begin{figure*}[!]
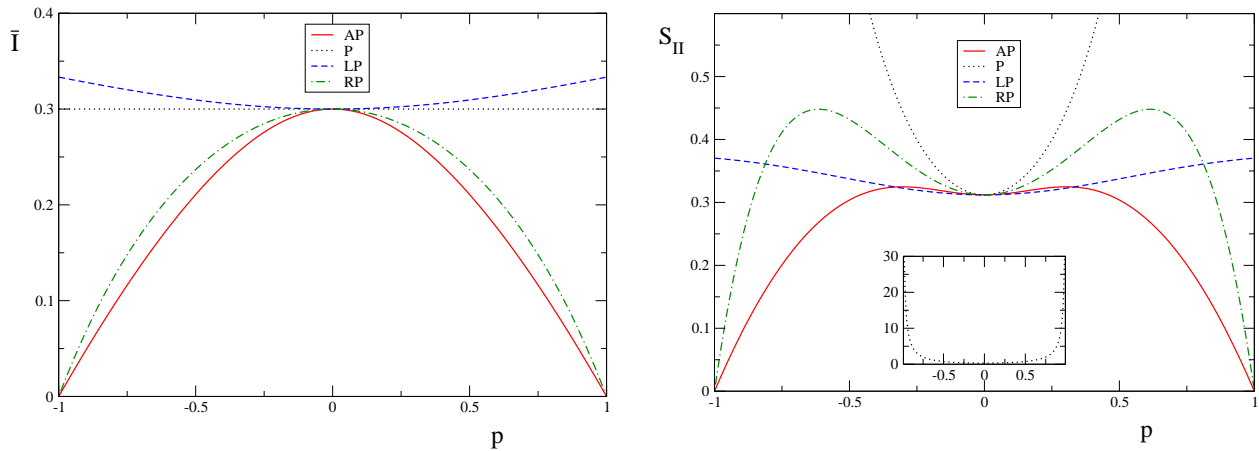

\begin{minipage}{8.5cm}
\includegraphics[width=8.cm]{fig3a.eps}
\end{minipage}
\begin{minipage}{8.5cm}
\includegraphics[width=8.cm]{fig3b.eps}
\end{minipage}
\caption{The mean current $\bar{I}$ (left panel) and its flucutations $S_{II}$ (right panel)
for the case of a molecular spin
$S=1/2$, 
for the four different cases of electrodes polarizations.  The inset in the right panel shows the
behavior of $S_{II}$ in the parallel case ($P$) on a larger scale.}
\label{fig:I12}
\end{figure*}

The behavior of the mean current as a function of the leads polarization,
for the case of the molecular spin $S=1/2$, is shown on the left panel of Fig.~\ref{fig:I12}.
The simplest case is the case of parallel polarizations (P, dotted curve): the mean current 
is then constant (value 3/10), and the polarization of the leads has thus no effect.
Note that the fact that the current is constant is related to the choice we made for
the density of states $\rho_{\uparrow/\downarrow,j} =   (1\pm p_j)/2 $ ($j=L,R$). Indeed,
we see that the total densitiy of states (spin up + spin down) in each electrode is constant.

Consider next the anti-parallel case (AP). We see (full curve on Fig.~\ref{fig:I12}) that the mean current has a maximum for
zero polarization, and decreases to zero when $p$ approaches $\pm 1$. This behavior can be
understood simply: when $|p|$ is large (let us take for example $p$ close to 1), the electrons
coming from the left electrode have preferentially a spin up, while the electrons going to the
right electrode have preferentially a spin down; the transport of such an electron from
the left to the right electrode implies thus a flip of the electron spin, and thus an increase of
1 of the $z$ component of the molecular spin. However, this will lead quickly to a maximally polarized molecular
spin, for which such a spin exchange will be impossible. The only processes contributing  to transport
will then involve the electrons with a low density of state (spin down in the left lead or spin up in the right one), 
for which the current goes to 0 when $|p|$ goes to 1. This behavior is a case of {\it spin blockade}: for $|p| = 1$,
an electron is blocked on the dot because its spin does not fit the collecting electrode spin.

The case where the right electrode only is polarized (RP, dash-dotted curve on Fig.~\ref{fig:I12}) is similar. 
For $p=1$, the system reaches a state where
the $z$ component of the molecular spin is maximally negative ($S_z=-S$ for the empty dot). In this state, spin-up
electron can tunnel from the left electrode to the right electrode without any splin-flip, but as soon as a spin-down electron
is tunneling from the left electrode, it is blocked on the dot because it cannot flip to a spin-up electron and tunnel to the right electrode.
Because of the very large Coulomb repulsion on the molecular level, the presence of this spin-down electron forbids any further
transport of spin-up electron. This case is thus also a case of spin blockade, as in the AP case, but the Coulomb repulsion on the molecular level
plays here a central role.
The decrease of the mean current with $p$ is a bit slower than in the AP case, as the process where an electron tunnel
without any spin flip, and without involving small densities of states, is always possible when $|p|<1$.
 
Finally, the case where the left electrode only is polarized (LP, dashed curve on Fig.~\ref{fig:I12}) 
 has a totally different behavior. There, the current is slowly increasing
when $|p|$ increases, and there is no spin blockade. Indeed, as the density of states of both spins are important in the right 
electrode, it is always possible to have transport of an electron without spin flip, and involving large densitites of states.
 
Let us now consider the zero-frequency current fluctuations, shown on the right panel of Fig.~\ref{fig:I12}. Again, we observe very
different behavior for the four different cases of electrode polarizations. At $p=0$, the value of $S_{II}$ is $39/125=0.312$.
In the parallel case (P, dotted curve), $S_{II}$ increases with $p$, and diverges as $(1-p^2)^{-1}$ as $|p|$ approaches 1 (see the inset in the figure).
 On the other hand, in the anti-parallel case (AP, full curve),
 the current fluctuations are nearly maximal at $p=0$ (with small shoulders near $|p|=0.3$), and decrease to 0 as $|p| \to 1$. 
This huge difference in behavior can be understood using the segment picture; let us take for example $p$ close to 1. 
In the parallel case, the most
probable process contributing to transport is simply the transport of one spin-up electron, without any spin flip, and thus without 
exchanging angular momentum with the molecular spin. The most probable segment is thus a very short one, with a single electron transfered.
However, an exchange of angular momentum (spin flip for the electron, and modification of the $z$ component of the molecular spin)
 can happen with a small probability; when this happen, the system will then transfer again a very large number of electrons without any spin flip, and it will take a very long time before the $z$ component of the molecular spin recover its inital value. There is thus a small probability
to have a very long segment, with many electrons transfered - the smaller the probability, the longer the segment. This presence of rare but 
arbitrary long segment when $p$ goes to 1, in a ``background'' of very short segments,
 explains the divergence of $S_{II}$ in the parallel case. The situation is different for the anti-parallel case:
for $p$ close to 1, the molecular spin is with a high probability in a maximaly polarized state ($S_z=S$ for the empty dot). The most
probable process is again the transfer of a single electron without spin flip (this produces a low current as it involves a small density
of state in one of 
the electrodes). There is again a small probability of a spin-flip, which will bring the molecular spin in the
state $S_z=S-1$. However, at this point, the most probable process (involving large densities of state in both electrodes)
tends to bring the molecular spin back to the $S_z=S$ 
state. The resulting segment is thus also short, with 2 (or at most a few) electrons tranferred.
As this probability for such longer segments goes to zero when $p \to 1$, and as the length of these segments is quite short, 
we understand why the current fluctuations $S_{II}$ go to 0 when $|p| \to 1$.

In the case where the right electrode only is polarized (RP, dash-dotted curve), 
the current fluctuations also go to 0 as $|p| \to 1$. Note however the presence
of broad shoulders, with a maximum of the fluctuations near $|p|=0.6$. Finally, in the case where the left electrode only is polarized (LP, dashed curve), the fluctuations have a behavior similar to the one of the mean current, with a slow increase when $|p|$ increases.

\subsection{$ \bar{J_z}$ and the $J_z$ fluctuations}

\begin{figure*}[!]
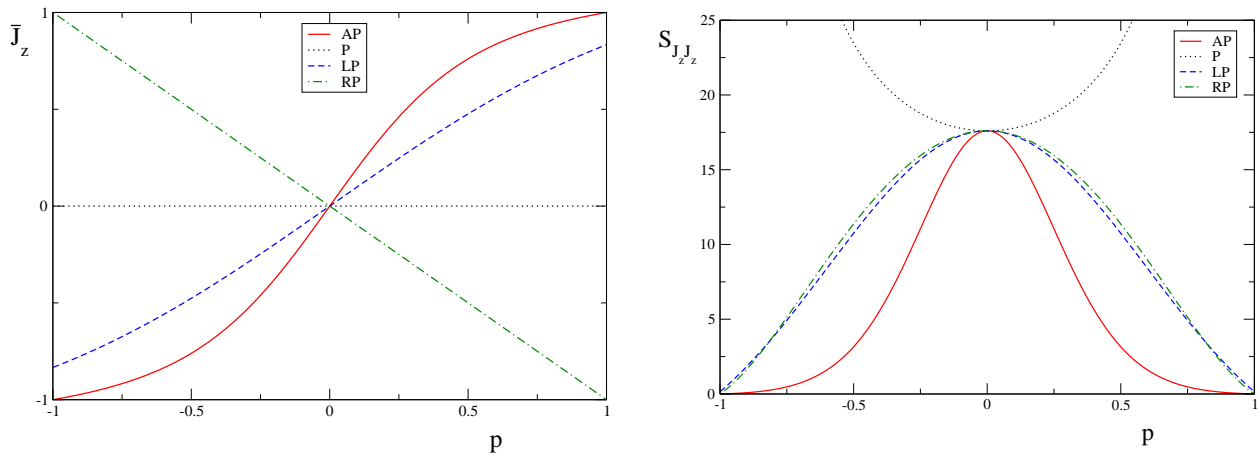

\begin{minipage}{8.5cm}
\includegraphics[width=8.cm]{fig4a.eps}
\end{minipage}
\begin{minipage}{8.5cm}
\includegraphics[width=8.cm]{fig4b.eps}
\end{minipage}
\caption{The mean value of the $z$ component of the molecular spin, $\bar{J_z}$ (left panel) and its flucutations $S_{J_zJ_z}$ (right panel)
for the case of a molecular spin 
$S=1/2$, 
for the four different cases of electrodes polarizations.}
\label{fig:Jz12}
\end{figure*}

In many works about electronic transport in the incoherent regime, the emphasis is put on the statistics of the electronic current, and little
attention is given to the statistics of other quantities (see however Ref. [\onlinecite{utsumi2007}]). Here  we study the statistics
of the total spin of the molecule (i.e. its $z$ component $J_z$), which gives us precious information on the impact of the electronic
current on the molecular spin.

The results for the mean value of the $z$ component of the molecular spin, $\bar{J_z}$, and the 
$J_z$ fluctuations $S_{J_zJ_z}$, are shown on Fig.~\ref{fig:Jz12}. Note that these two quantities involve both the molecular
spin when the dot is full ($J_z$) and the molecular spin for the empty dot (which is noted $S_z$, but as the intrinsic molecular
spin $S$ is also the total spin for an empty dot, $J_z$ reduces to $S_z$ for an empty dot).  The mean value $\bar{J_z}$ is an 
important quantity, as it shows how the current through the molecule is changing the polarization of its spin (as without any current,
one has simply $\bar{J_z} = 0$). The fluctuations $S_{J_zJ_z}$ show how the molecular spin fluctuates around its mean value,
it gives thus precious information on how precisely one could control the molecular spin polarization by applying a current.

In the parallel case (P, dotted curve on the left panel of Fig.~\ref{fig:Jz12}),
 the mean value $\bar{J_z}$ is simply 0 for all $p$. In
the anti-parallel case (AP, full curve), $\bar{J_z}$ is an non-linear odd function of $p$, 
going from 0 to 1 for $p$ going from 0 to 1. 
This behavior is easily understood. For $p>0$ for example, the spin-up electrons have a larger density of state than spin-down electrons in the left electrode, and it is the opposite in the right electrode. The system thus favors the transport 
of a spin-up electron from the left electrode into a spin-down electron in the right electrode 
(compared to the process with the spins exchanged), and this process increase $J_z$ by 1. 
On average $\bar{J_z}$ will thus be positive for $p>1$. For $p=1$,
we have seen that current is 0 because of spin blockade. In this case, the system is frozen in the state where the dot is full,
with $J_z=1$, hence one has $\bar{J_z}=1$.  

The behavior is quite similar in the case where the left electrode only is polarized (LP, dashed curve):
 $\bar{J_z}$ is an odd function of $p$,
positive for $p>1$. There are two main differences with the anti-parallel case. First, the slope at $p=0$ is smaller
 (it is 1 for LP, and 2 for AP). Next, $\bar{J_z}$ does not reach 1 (but 5/6) for $p=1$. This is because the current
is non-zero even for $p=1$ (no spin-blockade), and thus the molecular spin oscillates
 from the value $J_z=1$ (filled dot) and $S_z=1/2$ (empty dot).
 Finally, the case where the right electrode only is polarized (RP, dahs-dotted curve) has an opposite sign,
 and the behaviour is simply linear, with $\bar{J_z} = -p$. The fact that the sign is the opposite from the sign
 of the other cases ($\bar{J_z} = -1$ for $p=1$)
is due to the fact the tunneling of a spin-up electron to the right electrode (dominant for $p>1$) is done either without spin flip 
(if the electron coming from the left electrode is also a spin-up electron), or with a spin-flip which decreases $J_z$ by 1 (if the electron 
coming from the left electrode is a spin-down electron). The fact that $\bar{J_z}$ reaches the value $-1$ for $p=1$ 
is again due to spin blockade, as in the anti-parallel case.
 
Let us now consider the fluctuations of $J_z$ around its mean value, $S_{J_zJ_z}$ (right panel of Fig.~\ref{fig:Jz12}).
Note first that the value of $S_{J_zJ_z}$ for $p=0$ is $88/5=17.6$. This value, which determines the overall scale 
of the fluctuations, is extremely large, and is discussed in more details below.
In the parallel case (P), $S_{J_zJ_z}$ is increasing as $|p|$ increases, and has the same $(1-p^2)^{-1}$ divergence as the current noise
(for the same reasons). In the other three cases, $S_{J_zJ_z}$ is maximum at $p=0$, and decreases as $|p|$ increases, with a much
broader shapes for the cases with only one electrode polarized (LP and RP) compared to the anti-parallel case (AP). In the
anti-parallel case, and in the case where the right electrode only is polarized (RP), $S_{J_zJ_z}$ is 0 for $|p|=1$, because
of spin-blockade (current is zero). But in the case where the left electrode only is polarized (LP), $S_{J_zJ_z}=4/27$ for $|p|=1$  because
there is no-spin blockade: for $p=1$ the molecular spin oscillates between the values $S_z=1/2$ and $J_z=1$, leading
to these non-zero fluctuations. Finally, we note that the behavior of $S_{J_zJ_z}$ is very similar in the two 
cases where there is only
one electrode polarized (LP and RP); this is quite remarkable, as the currents (and the current noises) 
in these two cases have a completely different behavior (see Fig.~\ref{fig:I12}).

As said above, the scale of these $J_z$ fluctuations is very large, with a value $S_{J_zJ_z} = 88/5$ for $p=0$. A natural normalization
of these fluctuations, to take into account the value of the molecular spin, is to divide by $(2S+1)^2$ 
(note that $\hbar=1$). Here, 
$S=1/2$, which gives a normalized value of $22/5$. 
This value has the dimension of a time, and should be compared with a typical
time in the system. Here, the natural time is just the inverse transtion rate $1/\Gamma^{(0)}_{L,R}=1$, 
which gives the scale of the time
to transfer an electron. The value of the $J_z$ fluctuations at $p=0$ is quite larger than this time scale. 
Comparison with the value for the charge fluctuations $S_{QQ}$, which is $96/125 \simeq 0.77$ (see table~\ref{table:ana12}) shows also that the normalized $J_z$ fluctuations for $p=0$ are extremey large. Even if these fluctuations decrease with increasing $|p|$ (except in the parallel case),
 they remain quite large when $|p|$ is not close to 1.
 One can thus speak of {\it colossal spin fluctuations}, and this implies 
that it is difficult to control the molecular spin with 
the current, except with electrodes having polarizations $p$ close to 1.

In this respect, the anti-parallel case is  much more favourable than the case where only one electrode is polarized.
One could for example think to use a setup with only one polarized electrode,
to flip the molecular spin by reversing the current in the setup. Indeed, reversing the voltage bias will make the system go
from the LP case to the RP case. If $p=0.5$ for the polarized electrode, we see on the left panel of Fig.~\ref{fig:Jz12}
that $\bar{J_z}$ would change from approximatively $0.5$ (case LP) to $-0.5$ (case RP) when the bias voltage is reversed. 
However, as the $J_z$ fluctuations are very large (the normalized value is $\simeq 5$), it is difficult to say that the molecular
spin is {\it controlled}. Performing the same with two polarized electrodes in the anti-parallel polarization configuration 
would be more effective: reversing the bias voltage is then equivalent to the change $p \to -p$, and with $p=0.5$ it would change 
$\langle J_z \rangle$ from approximatively $0.75$ to $-0.75$. The normalized value of
the $J_z$ fluctuations is then approximatively $1.3$, which is much lower than in the previous case.

\section{Behavior for larger molecular spin}
In the previous section, we have shown the results obtained in the case of a molecular spin $S=1/2$. The method we have
presented is of course not limited to this value of the spin, and we discuss in this section how the results are changed when the molecular spin is larger than $1/2$. Note that the analytical results for $S=1$ are given in table~\ref{table:ana1e}, in appendix~\ref{app:tableana1}. The plots
one obtains with these results are qualitatively similar to the ones for $S=1/2$.

\begin{figure*}[!]
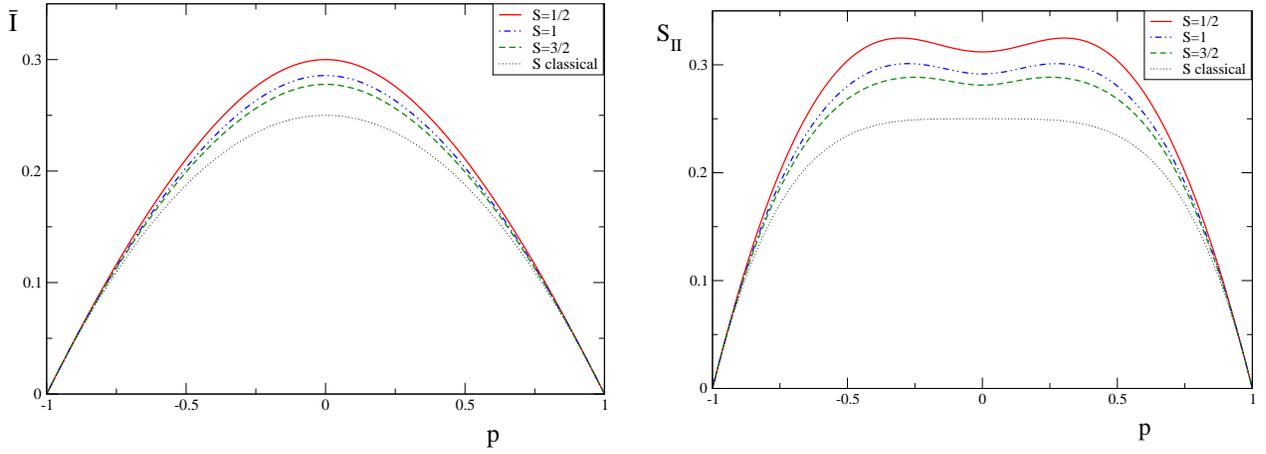

\begin{minipage}{8.5cm}
\includegraphics[width=8.cm]{fig5a.eps}
\end{minipage}
\begin{minipage}{8.5cm}
\includegraphics[width=8.cm]{fig5b.eps}
\end{minipage}
\caption{The mean current $\langle I \rangle$ (left panel) and its fluctuations $S_{II}$ (right panel), in the
anti-parallel configuration of the electrodes, for the values of the spin $S=1/2$, $S=1$ and $S=3/2$ as 
a function of the electrodes polarizations $p$. When increasing the spin $S$, the curves converge towards
a ``classical'' curve (dotted line) corresponding to a fixed spin.}
\label{fig:IlargeS}
\end{figure*}

In order to discuss the behavior at larger $S$, one should distinguish between the mean current (and the current fluctuations) on the one hand, and the
mean value of $J_z$ (and the $J_z$ fluctuations) on the other hand. For the mean current, and the current fluctuations, there is very little
change as one increases the spin, and the physical explanations we have given for $S=1/2$ apply for arbitrary spin. This is illustrated
on Fig.~\ref{fig:IlargeS}, which shows the mean current $\bar{I}$ (left panel) and the current fluctuations $S_{II}$ (right panel)
 as a function of the polarization $p$, for the anti-parallel (AP) configuration, and for the values of the spin $S=1/2$, $S=1$ and $S=3/2$.
One can see that both the mean current and the current fluctuations decrease a little bit as the spin is
increased, with very little change in the $p$ dependence. For larger values of $S$, the curves will slowly converge towards a 
``classical curve'', obtained by considering a classical (fixed) spin.

\begin{figure*}[!]
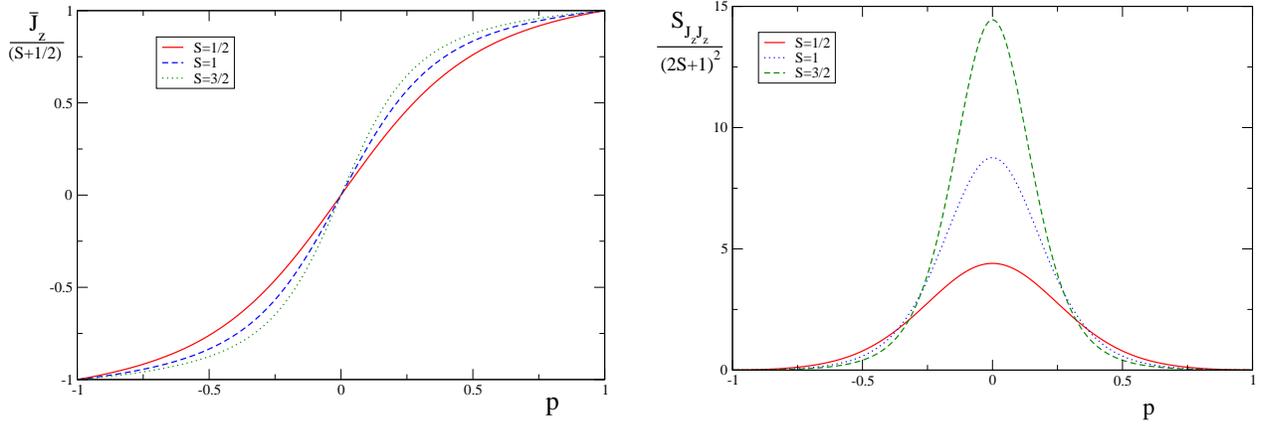

\begin{minipage}{8.5cm}
\includegraphics[width=8.cm]{fig6a.eps}
\end{minipage}
\begin{minipage}{8.5cm}
\includegraphics[width=8.cm]{fig6b.eps}
\end{minipage}
\caption{The normalized mean value of $J_z$, $\langle J_z \rangle/(S+1/2)$ (left panel) and its normalized fluctuations $S_{J_zJ_z}/(2S+1)^2$
 (right panel), in the anti-parallel configuration of the electrodes, for the values of the spin $S=1/2$, $S=1$ and $S=3/2$ as 
a function of the electrodes polarizations $p$.}
\label{fig:JzlargeS}
\end{figure*}

The situation for the mean of $J_z$ and its fluctuations is slightly
different. First, as $J=S+1/2$, it is natural to normalize the results 
for the different $S$ to compare them; we normalize $\bar{J_z}$ by $S+1/2$, and the $J_z$ fluctuations by $(2S+1)^2$. The normalized
mean of $J_z$ is shown on the left panel of Fig~\ref{fig:JzlargeS}, for the case of anti-parallel polarizations (AP), and for the
values of the spin $S=1/2$, $S=1$ and $S=3/2$. When $p \to 1$, the molecular spin is maximaly polarized, and thus each curve reaches the value
1 for $p=1$. However, the slope at $p=0$ increases when $S$ increases. By inspection of the formulas, we see that the slope at $p=0$ is given
by $4/3 * (S+1)$. The molecular spin is thus more easily polarized when $S$ increases. When the molecular spin becomes large,
we expect that it becomes more sensitive to the electrodes magnetizations, reaching even for small $|p|$ highly polarized
states.

 The $J_z$ fluctuations, $S_{J_zJ_z}$, normalized
by $(2S+1)^2$, are shown on the right panel of Fig~\ref{fig:JzlargeS}. Two important characteristics appear on this figure.
First, the maximum value of the normalized fluctuations, for $p=0$, increase strongly when $S$ increase. This maximum is already very
large for $S=1/2$ (see the discussion in the previous section), but it is still much larger for larger $S$.
At $p=0$, $S_{J_zJ_z}=88/5$ for $S=1/2$,
$S_{J_zJ_z}=552/7$ for $S=1$, and
$S_{J_zJ_z}=2080/9$ for $S=3/2$, $\cdots$.
Secondly, the width of the curves as a function of $p$ decreases as $S$
increases: the full width at half maximum is $\simeq 0.6$ for $S=1/2$, $\simeq 0.45$ for S=1 and $\simeq 0.35$ for $S=3/2$. This means that 
for $|p|$ large enough, the normalized fluctuations decrease when $S$ is increased. 
For example, on the figure, we see that for $|p|\gtrsim 0.4$,
the normalized $J_z$ fluctuations for $S=3/2$ are smaller than those for $S=1/2$. 
We expect this tendency to continue when $S$ is increased, with the normalized $J_z$ fluctuations for a large spin $S$ having 
the shape of a narrow peak with a very large maximum value. As soon as the electrodes have some magnetization, a larger spin is thus  relatively
easier to control than a small spin $S$, as the normalized $J_z$ fluctuations can be much smaller.

\section{Alternative method of calculation}
We discuss here shorlty another method of calculation, which can also be used to obtain analytical formulas for the averages and fluctuations of different
quantities, and which can even give access to analytic expression for the higher moments. 
It is derived from the master equation approach to the full counting
statistics, which was introduced in Ref.[\onlinecite{bagrets_nazarov}], and adapted for a molecular quantum dot magnet (placed between normal electrodes)
in Ref.[\onlinecite{fcs2007}]. A full explanation of the method can be found in these two references.  

In the master equation approach, a $n \times n$ matrix $L$ determines the time evolution of the populations of the $n$ different states of the system :
\begin{equation}
\frac{d p(t)}{dt} = L \; p(t)
\end{equation}
where $p(t)$ is the vector containing the populations. Any off-diagonal element $L_{ij}$ of the matrix $L$ gives the transition probability from state
$i$ to state $j$. To obtain the full counting statistics of a given physical quantity, a counting field $\xi$ is introduced in the matrix $L$, by making the appropriate replacements of diagonal and non-diagonal elements. The eigenvalues of the matrix $L(\xi)$ then give
access to the full counting statistics, as the cumulant generating function is simply proportionnal to the eigenvalue $\lambda(\xi)$ which satisfies
$\lim_{\xi \to 0} \lambda(\xi) = 0$. 
The full counting statistics is thus obtained by solving the equation:
\begin{equation}
\lambda^n + f_{n-1}(\xi) \lambda^{n-1} + \dots + f_1(\xi) \lambda + f_{0}(\xi) = 0 ,
\label{eq:caracpoly}
\end{equation}
where the functions $f_{i}(\xi)$ ($i=0,\dots,n-1$) depend on the matrix $L$ modified by the counting field $\xi$. 
The solution $\lambda(\xi)$ which satisfies $\lim_{\xi \to 0} \lambda(\xi) =0$ then gives access to the cumulants $C_n(X)$
of the quantity $X$ associated with the counting field $\xi$:
\begin{equation}
C_n(X) = \left.\frac{\partial^n \lambda(\xi)}{\partial \xi^n}\right|_{\xi=0} .
\label{eq:lambda}
\end{equation}
 Note that the first two cumulants ($C_1$ and $C_2$) are simply the average and the fluctuations which have been calculated in the previous
sections. In general, the cumulant of order $n$ can be expressed as a combination of the centered moments of order $\leq n$.\cite{Belzig} $T$ is the measuring time, which must be larger than all typical times in the system.
It is in general impossible to solve Eq.~(\ref{eq:caracpoly}) analytically 
(except in special cases which can be reduced to small $n$, as in Ref.[\onlinecite{fcs2007}]).
 However, if one is interested in the cumulants up to a finite order $n_{max}$ only,
then one can expand the function $\lambda(\xi)$ in power of $\xi$, and keep only the terms up to the order $n_{max}$. 
It is then possible to solve Eq.~(\ref{eq:caracpoly}) by expanding all the terms in powers of $\xi$, and by solving order by order,
starting from order 1, up to order $n_{max}$. Specifically, to compute the average and fluctuations (of the chosen quantity) only, it is 
enough to write $\lambda(\xi) = C_1 \xi + (C_2/2) \xi^2$, to develop $f_0(\xi)$, $f_1(\xi)$ and $f_2(\xi)$ up to order 2 in $\xi$, and
then to solve Eq.~(\ref{eq:caracpoly}) first for $C_1$ (terms in $\xi$) and then for $C_2$ (terms in $\xi_2$).   

There are two kinds of observable with different types of countings fields. First, charge-like operators, which have a given value
for each state of the dot (for example the charge $Q$, or $J_z$). In this case, the counting field is simply introduced by adding 
$+ c_{\alpha} \xi$ to each diagonal element $L_{\alpha\alpha}$, where $c_{\alpha}$ is the value of the observable (for example $J_z$) in state $\alpha$. Secondly,
current-like operators, which are associated with transitions between different states (the charge current being the main example). In this
case, the counting field is introduced by multiplying the off-diagonal elements $L_{\alpha\beta}$ which are associated to transitions contributing to the 
current by $e^{i \xi}$.  For this second kind of observable,
there is an additional factor $i^n$ on the right hand side of Eq.(\ref{eq:lambda}).

\begin{figure*}[!]
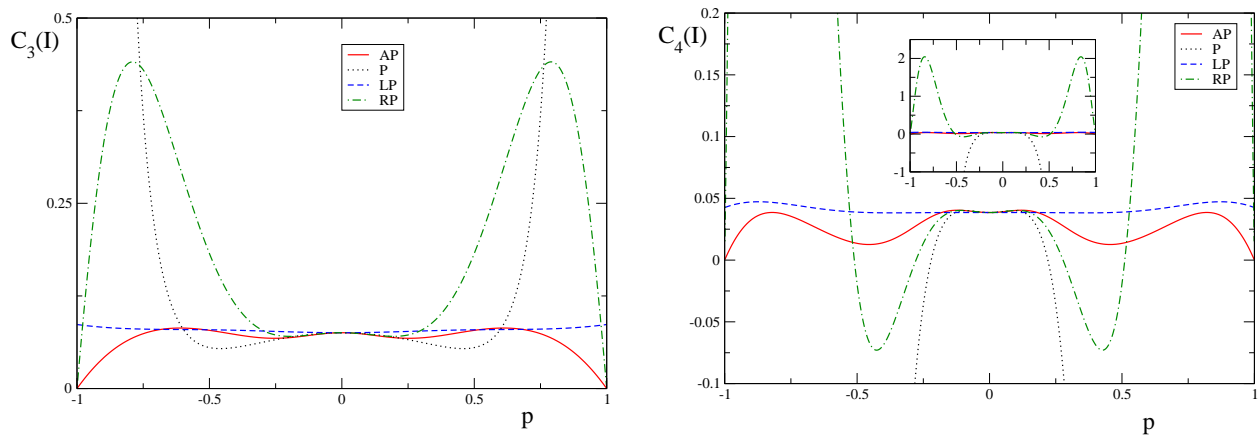

\begin{minipage}{8.5cm}
\includegraphics[width=8.cm]{fig7a.eps}
\end{minipage}
\begin{minipage}{8.5cm}
\includegraphics[width=8.cm]{fig7b.eps}
\end{minipage}
\caption{The third cumulant of the the current $C_3(I)$ (left), and the fourth cumulant of the current $C_4(I)$ (right) as
a function of the electrodes polarizations $p$, for the four different cases of electrodes polarizations. The inset on the right panel shows
the curves on a larger scale.}
\label{fig:Icumulants}
\end{figure*}

With this alternative method, we have computed all the quantities shown in the previous sections of this article, 
and we verified that we could indeed recover 
the same formulas. 
This method allow to compute quite easily
cumulants of order higher than 2. We do not provide here a complete exploration of the higher cumulants of the physical
quantities we are interested in, but as an example Fig.\ref{fig:Icumulants} shows the third and fourth cumulant of the current for the
four possible cases of electrode magnetizations.
We see on these plots that some of the features present for the current fluctuations (Fig.\ref{fig:I12}) are more pronounced on the third and
fourth cumulant. In the right-lead polarized case (RP), peaks for large $p$ are becoming more pronounced, while in the left lead polarized case (LP), the 
variations of the cumulant as a function of $p$ are extremely small.

\section{Conclusion}
In this paper, we have studied the incoherent transport between ferromagnetic electrodes through a magnetic molecule with an isotropic spin.
 The molecule is modeled as a single 
resonant level, with large electronic interaction forbidding double occupancy of the level. There is an exchange coupling between the molecular
spin and the electronic spin on the molecular level. By extending an original method (which was first introduced by Korotkov in the study of noise 
of the singe-electron transistor~\cite{korotkov1994}), we have shown that it is possible to obtain analytical formulas for the average and fluctuations
of all physical quantities of interest. The idea of the method is to separate the transport
process in statistically independent segments, and to compute the average and fluctutations using the properties of individual
segments and then averaging using the statistical distribution of all segments.
 We have focused on the current through the molecule and its fluctuations, and on the total spin of the molecule and its fluctuations.
 We limited the calculations to the case of a temperature
$\Theta$ much smaller than the bias voltage, and with only the level $J=S+1/2$ in the bias window, but using the same method it would be quite easy 
to generalize the results to higher temperatures, or to the case where the two levels $J=S \pm 1/2$ are inside the bias window.

 By considering several configurations of the magnetization of the electrodes (both electrodes polarized with parallel or anti-parallel polarizations, 
or only one electrode polarized), we have shown that there is a rich variety of behaviors. Depending on the electrode polarizations,
the current can decrease or increase with the electrodes polarization; for maximally polarized electrodes, it can be blocked due to spin blockade.
The current fluctuations can show a non-monotonic behavior as polarization is increased.

With the study of $J_z$ (the $z$ component of the total spin of the molecule), we have been able to characterize how the electronic transport 
affects the molecular spin.  If the results for the average of $J_z$ show, as expected, that it is possible to polarize the molecular spin by using magnetic
electrodes (non-zero average of $J_z$ whose sign depends on the sign of the current),
 the results for the fluctuations of $J_z$ show that these fluctuations are very large. This shows that it is effectively
difficult to control the molecular spin with the current, except with electrodes having polarizations close enough to 1. We have
shown how these results evolve when increasing the bare molecular spin: the polarizability of the molecular spin increases near $p=0$,
and the fluctuations of $J_z$ are more peaked around $p=0$. These large fluctuations of the molecular spin are of course a direct 
consequence of the isotropy of the molecular spin. The fluctuations would be severly reduced in molecular magnets, where a strong spin-anisotropy
is present. 

We have also shown that we can obtain the same results using a different calculation, based on an extension of the method introduced by Bagrets 
and Nazarov~\cite{bagrets_nazarov} to compute the full counting statitics in Coulomb blockade systems. In contrast to the segment method,
this second method offers the possibility to compute the higher cumulants of the physical quantities, 
and we have shown as an example the results for the third and fourth cumulant of the electronic current. This second method gives however
less information for the physical interpretation of the results, compared to the segment method.

\vspace{3.cm}
\appendix

\section{Explicit formulas I: First average --- an average in a given segment} 
In the segment picture, the statistical average $\bar{X}$ and the fluctuations 
$S_{XX}$ of a given random variable $X(t)$,
reduces to an average of quantities {\it in a single segment}, such as,
$\langle{\cal X}\rangle, \langle\tau\rangle, \langle{\cal X}\tau\rangle,\cdots$,
and its lowest order examples are, Eqs. (\ref{Xav},\ref{Xnoise}).
In these formulas, the average in a single segment is done in two steps,
i.e., first a Poissonian average over a given segments is taken, and then
one takes an average over different segments with suitable weights.
One may denote the first average for a given segment $\zeta$ as
$\langle\cdots\rangle_1$. 
The probability with which in a given segment
$\zeta=(\alpha_1,\alpha_2,\cdots,\alpha_{M-1})$,
the $m$-th state happens to survive during a period $\tau_m$
is proportional to $\exp[-\Gamma_m\tau_m]$,
where $1/\Gamma_m$
is the average lifetime of the $m$-th state.
The probability with which $\zeta$ is characterized by a set of transition
times $(\tau_0,\tau_1,\cdots,\tau_{M-1})$ thus obeys to a Poissonian distribution,
\begin{equation}
P_1(\tau_0,\tau_1,\cdots,\tau_{M-1})=\Gamma_0 \Gamma_1 \cdots\Gamma_{M-1}
e^{-\Gamma_0 \tau_0}e^{-\Gamma_1 \tau_1}\cdots e^{-\Gamma_{M-1} \tau_{M-1}},
\end{equation}
and in that case, one can rewrite the probability $P[\zeta]$ of the segment $\zeta$ as
$P[\zeta]=P_1(\tau_0,\tau_1,\cdots,\tau_{M-1})P_2[\zeta]$.
Then, together with the explicit definition of the first average, 
\begin{equation}
\langle{\cal X} [\zeta] \rangle_1=
\int d\tau_0 d\tau_1\cdots d\tau_{M-1}
{\cal X} [\zeta] P_1(\tau_0,\tau_1,\cdots,\tau_{M-1}),
\label{av1}
\end{equation}
one indeed arrives at Eq. (\ref{av12}).
In Eq. (\ref{av1}), ${\cal X} [\zeta]$ is, e.g., ${\cal J}_z[\zeta]$ defined as Eq. (\ref{jzxi}).
As for the second average, Section II B demonstrates how to construct explicitly
$P_2[\zeta]$ in the case of molecular quantum dot magnet.

At linear order, the role of $\langle\cdots \rangle_1$ is nothing more than
a replacement $\tau_m\rightarrow \langle\tau_m \rangle_1=1/\Gamma_m$,
e.g.,
\begin{equation}
\langle{\cal J}_z[\zeta]\rangle_1=
\sum_{m=0}^{M-1}J^z_m \langle\tau_m\rangle_1=
\sum_{m=0}^{M-1}{J^z_m \over \Gamma_m}.
\end{equation}
At second order, one may also use the following mathematical trick \cite{korotkov1994}
Let us attempt to calculate, e.g.,
$\langle{\cal J}_z[\zeta]\tau[\zeta]\rangle_1$:
\begin{eqnarray}
\langle{\cal J}_z[\zeta]\tau[\zeta]\rangle_1&=&
\left\langle
\sum_{m=0}^{M-1} J^z_m \tau_m
\sum_{m'=0}^{M-1} \tau_{m'}
\right\rangle_1=
\left\langle
\sum_{m=0}^{M-1} J^z_m(\tau_m)^2+
\sum_{m\neq m'} J^z_m\tau_m\tau_{m'}
\right\rangle_1
\nonumber \\
&=&
\sum_{m=0}^{M-1} J^z_m \langle(\tau_m)^2\rangle_1+
\sum_{m\neq m'} J^z_m 
\langle\tau_m\tau_{m'}\rangle_1
\nonumber \\
&=&
\sum_{m=0}^{M-1} J^z_m {2\over\Gamma_m^2}+
\sum_{m\neq m'} J^z_m 
{1\over\Gamma_m}{1\over\Gamma_{m'}}
\nonumber \\
&=&
\sum_{m=0}^{M-1} {J^z_m\over\Gamma_m^2}+
\langle{\cal J}_z[\zeta]\rangle_1 \langle\tau[\zeta]\rangle_1
\label{trick}
\end{eqnarray}
At the second identity, we divided the double summation into diagonal 
and off-diagonal parts.
To give some concrete examples, 
in the case of segments $\zeta=A$ and $\zeta=B^*(i_1,i_2,\cdots,i_{4S})$ 
defined in Section II,
the final expression reads explicitly as,
\begin{eqnarray}
&&\langle{\cal J}_z[A]\tau [A]\rangle_1
=
{S\over \big[\Gamma_{Q=0}(S_z=S)\big]^2}+
{S+1/2\over \big[\Gamma_{Q=1}(J_z=S+1/2)\big]^2}+
\langle{\cal J}_z[A]\rangle_1 \langle\tau [A]\rangle_1,
\nonumber \\
&&\langle{\cal J}_z[B^*]\tau[B^*]\rangle_1=
{S\over \big[\Gamma^{Q=0}_{S_z=S}\big]^2}+
{S-1/2\over \big[\Gamma^{Q=1}_{S_z=S-1/2}\big]^2}
\label{jztauA}
\\
&+&\sum_{l=1}^{2s}i_{2l-1}
\left(
{S-l+1/2\over \big[\Gamma^{Q=1}_{J_z=S-l+1/2}\big]^2}+
{S-l\over \big[\Gamma^{Q=0}_{S_z=S-l}\big]^2}
\right)
+\sum_{l=1}^{2s}i_{2l}
\left(
{S-l\over \big[\Gamma^{Q=0}_{S_z=S-l}\big]^2}+
{S-l-1/2\over \big[\Gamma^{Q=1}_{J_z=S-l-1/2}\big]^2}
\right)
\nonumber \\
&+&\langle{\cal J}_z[B^*]\rangle_1
\langle\tau [B^*] \rangle_1,
\label{jztauB}
\end{eqnarray}
where we used an abbreviated notation,
$\Gamma^{Q=0}_{S_z}=\Gamma_{Q=0}(S_z)=\Gamma_{Q=0}^+(S_z)+\Gamma_{Q=0}^-(S_z)$, and
$\Gamma^{Q=1}_{J_z}=\Gamma_{Q=1}(J_z)=\Gamma_{Q=1}^+(J_z)+\Gamma_{Q=1}^-(J_z)$.
Substituting Eqs. (\ref{jztauA},\ref{jztauB}) into Eq. (\ref{av12}) in Section II A, 
and together with $P_2[A]$ and $P_2[B^*]$ constructed in Section II B (Eqs. (\ref{probA},\ref{probB})), 
one finds $\langle{\cal J}_z\tau \rangle$ appearing in Eq. (\ref{Xnoise}) for $X=J_z$. 
Evaluating other averages in a single segment, such as 
$\langle\tau \rangle$, $\langle{\cal J}_z\rangle$, $\langle\tau^2 \rangle$, and $\langle{\cal J}_z^2 \rangle$,
one finally finds the expression for $S_{J_zJ_z}=2\mu_2[J_z]$.

\section{Explicit formulas II: Second average --- average over different segments and consistency with the FCS generating function} 
To illustrate how to evaluate the average over different segments, let us give here
some explicit formulas, which typically appear in the calculation.
For simplicity, we consider here only the case of $S=1/2$, and 
parallel (P) or anti-parallel (AP) spin alignment of the electrodes: 
$P_L=P_R=p$ for P, and  $P_L=-P_R=p$ for AP.
Using Eq. (\ref{av12}) in Section II A, and some explicit formulas in Appendix A and
in Section II B (expressions for $P_2[A]$ and $P_2[B^*]$, in particular, i.e., Eqs. (\ref{probA},\ref{probB})), 
one finds the average duration of a segment as,
\begin{eqnarray}
\langle\tau\rangle_P&=&{2\over 3+p}{3\Gamma_L+2\Gamma_R\over\Gamma_L\Gamma_R},
\nonumber \\
\langle\tau\rangle_{AP}&=&
{2(3+10p^2+3p^4)\Gamma_L+4(1-p^4)\Gamma_R
\over (1+p)^3(3-2p-p^2)\Gamma_L\Gamma_R}.
\label{tau}
\end{eqnarray}
Here, we used a slightly different convention from the body of the paper,
so that we can compare our results directly with that of Ref. [\onlinecite{fcs2007}]. 
The convention here is
$\Gamma_{L,R}^\uparrow=\Gamma_L(1+p_{L,R})$ and
$\Gamma_{L,R}^\downarrow=\Gamma_L(1-p_{L,R})$.
Note that Eqs. (\ref{tau}) are not symmetric functions of $p$, and this reflects the choice of the reference state.
The average net charge on the dot is,
\begin{eqnarray}
\langle{\cal Q}\rangle_P={6\over 3+p},\ \ \
\langle{\cal Q}\rangle_{AP}=
{2(3+10p^2+3p^4)\over (1+p)^3(3-2p-p^2)\Gamma_R},
\end{eqnarray}
no longer a symmetric function of $p$.
On the other hand, the average charge normalized by the average duration of a segment, which is the physically measurable average charge,
is a symmetric function of $p$, indepedent of the choice of the reference state:
\begin{eqnarray}
\bar{Q}_P&=&{\langle {\cal Q}\rangle_P\over\langle\tau\rangle_P}
={3\Gamma_L\over 3\Gamma_L+2\Gamma_R},
\nonumber \\
\bar{Q}_{AP}&=&{\langle {\cal Q}\rangle_{AP}\over\langle\tau\rangle_{AP}}
={(3+10p^2+3p^4)\Gamma_L\over (3+10p^2+3p^4)\Gamma_L+2(1-p^4)\Gamma_R}.
\label{qav}
\end{eqnarray}
As for the current, one has to evaluate similarly,
$\langle k\rangle_P$ or $\langle k\rangle_{AP}$
to find,
\begin{eqnarray}
\bar{I}_P&=&{\langle k\rangle_P\over\langle\tau\rangle_P}
={3\Gamma_L\Gamma_R\over 3\Gamma_L+2\Gamma_R},
\nonumber \\
\bar{I}_{AP}&=&{\langle k\rangle_{AP}\over\langle\tau\rangle_{AP}}
={(1-p^2)(3+5p^2)\Gamma_L\Gamma_R\over (3+10p^2+3p^4)\Gamma_L+2(1-p^4)\Gamma_R}.
\label{iav}
\end{eqnarray}
Here, we used the same notation $k$ as Ref. \cite{korotkov1994} to ease the comparison.
Observable quantities, i.e., the charge, current, or spin 
{\it averaged over measurement time},
are either a symmetric (even) or an antisymmetric (odd)
function of $p$ depending on their symmetry properties under 
spin reversal.
One can verify, for example, that $J_z$ averaged over the measurement time is an odd function 
of $p$. 
\begin{eqnarray}
\bar{J_z}_{P}&=&0,
\nonumber \\
\bar{J_z}_{AP}&=&{\langle {\cal J}_z\rangle_{AP} \over\langle\tau\rangle_{AP}}
={2p\{4(1+p^2)\Gamma_L+(1-p^2)\Gamma_R\}\over (3+10p^2+3p^4)\Gamma_L+2(1-p^4)\Gamma_R}.
\label{jzav} 
\end{eqnarray}
Note that the average $J_z$ vanishes for P-alignment.

At second order, one finds expressions, such as ($\mu_2$ is the centered moment of second order)
\begin{eqnarray}
\mu_2[Q]_P&=&
{24\Gamma_L\Gamma_R
\over (1-p^2)(3\Gamma_L+2\Gamma_R)^3},
\nonumber \\
\mu_2[Q]_{AP}&=&
{8(1-p^2)(3+32p^2+38p^4+40p^6+15p^8)\Gamma_L\Gamma_R
\over
\{(3+10p^2+3p^4)\Gamma_L+2(1-p^4)\Gamma_R\}^3}.
\label{q2av}
\end{eqnarray}
Note the these correlation functions are actually cumulants or
fluctuations around the mean value.
At this order, expressions start to be lengthy, so that we list here only a few examples of
our results:
\begin{eqnarray}
\mu_2[I]_P&=&
{\Gamma_L\Gamma_R\{27(1+3p^2)\Gamma_L^2+48p^2\Gamma_L\Gamma_R
+4(3+p^2)\Gamma_R^2
\over
(1-p^2)(3\Gamma_L+2\Gamma_R)^3}
\label{i2av} 
\\
\mu_2[J_z]_{AP}&=&
{2(6\Gamma_L^2+4\Gamma_L\Gamma_R+\Gamma_R^2)
\over
(1-p^2)\Gamma_L\Gamma_R(3\Gamma_L+2\Gamma_R)}
\label{jz2av} 
\end{eqnarray}

In order to check the consistency of these results,
let us compare them with the FCS generating function $\Omega(\xi,\eta)$.
For non-magnetic electrodes ($p=0$), the analytic expression for $\Omega(\xi,\eta)$ 
is given in Ref. \cite{fcs2007} as,\cite{erratum}
\begin{equation}
\Omega(\xi,\eta)=T\left[{\xi-z\Gamma_L-\Gamma_R\over 2}+{1\over 2}
\sqrt{(z\Gamma_L-\Gamma_R+\xi)^2+4z\Gamma_L\Gamma_R e^\eta}
\right],
\label{cgf}
\end{equation}
where $z=(2S+2)/(2S+1)$, i.e., $z=3/2$ for $S=1/2$.
Taking derivatives of Eq. (\ref{cgf}) with respect to counting fields $\xi$ or $\eta$,
one can, in principle, obtain any correlation function associated with $Q$ and $I$,
i.e.,
\begin{equation}
\kappa_{m,n}[Q,I]=\left. {1\over T}\langle\langle Q^m I^n \rangle\rangle_c\right|_{p=0}=
\left.{1\over T}{\partial^m \over \partial \xi^m} {\partial^n \over\partial\eta^n} 
\Omega(\xi,\eta)\right|_{\xi\rightarrow 0, \eta\rightarrow 0}.
\end{equation}
At lowest orders, this gives,
\begin{eqnarray}
\kappa_1 [Q]_{p=0}&=&\bar{Q}_{p=0}=
{3\Gamma_L\over 3\Gamma_L+2\Gamma_R},\ \ \
\kappa_1 [I]_{p=0}=\bar{I}_{p=0}={3\Gamma_L\Gamma_R\over 3\Gamma_L+2\Gamma_R}
\nonumber \\
\kappa_2 [Q]_{p=0}&=&\mu_2 [Q]_{p=0}=
{24\Gamma_L\Gamma_R
\over (3\Gamma_L+2\Gamma_R)^3},\ \ \
\kappa_2 [I]_{p=0}=\mu_2 [I]_{p=0}=
{3\Gamma_L\Gamma_R\ (9\Gamma_L^2+4\Gamma_R^2)
\over (3\Gamma_L+2\Gamma_R)^3}.
\label{check}
\end{eqnarray}
One can, therefore, check the consistency between Eqs. (\ref{qav},\ref{iav},\ref{q2av},\ref{i2av}) 
and Eq. (\ref{cgf}), by verifying the formulas given in Eqs. (\ref{check}).
The formulas (\ref{qav},\ref{iav},\ref{jzav},\ref{q2av},\ref{i2av},\ref{jz2av}) are also listed in Table I
in the limit of $\Gamma_L\rightarrow 1$ and $\Gamma_R\rightarrow 1$.

\section{Table for $S=1$}
\label{app:tableana1}
\begin{table*}
\[
\begin{array}{|c||c|c|}
\hline
P &  \langle I \rangle & \displaystyle{ 2/7}   \\ \hline
AP &  \langle I \rangle & \displaystyle{ -\frac{2 (p-1) (p+1) \left(7 p^4+14 p^2+3\right)}{3
   \left(p^6+21 p^4+35 p^2+7\right)} }   \\\hline
LP &  \langle I \rangle & \displaystyle{ \frac{2 \left(p^2+1\right)}{5 p^2+7}}\\\hline
RP &  \langle I \rangle & \displaystyle{ \frac{2 \left(p^4+2 p^2-3\right)}{3 \left(p^4-2
   p^2-7\right)} }  \\\hline\hline 
P &  S_{II} & \displaystyle{-\frac{4 \left(539 p^2+75\right)}{1029
   \left(p^2-1\right)}  } \\\hline
AP &  S_{II} & \displaystyle{\frac{4 (p-1) (p+1) \left(357 p^{16}-2698 p^{14}-24654
   p^{12}-71634 p^{10}-100500 p^8-68254 p^6-23458
   p^4-3846 p^2-225\right)}{9 \left(p^6+21 p^4+35
   p^2+7\right)^3} }  \\\hline
LP &  S_{II} & \displaystyle{\frac{4 \left(27 p^6+27 p^4+81 p^2+25\right)}{\left(5
   p^2+7\right)^3} } \\\hline
RP &  S_{II} & \displaystyle{\frac{4 (p-1) (p+1) \left(3 p^{10}+85 p^8+270 p^6-518
   p^4+1215 p^2+225\right)}{9 \left(p^4-2 p^2-7\right)^3} }\\\hline\hline 
P & \langle J_z \rangle & \displaystyle{ 0 } \\ \hline
AP & \langle J_z \rangle & \displaystyle{ \frac{4 p \left(3 p^4+14 p^2+7\right)}{p^6+21 p^4+35
   p^2+7} } \\ \hline
LP & \langle J_z \rangle & \displaystyle{ \frac{2 p \left(p^2+7\right)}{5 p^2+7} } \\ \hline
RP & \langle J_z \rangle & \displaystyle{ -\frac{2 p \left(p^2-7\right)}{p^4-2 p^2-7} } \\ \hline \hline
P & S_{J_zJ_z} & \displaystyle{ \frac{552}{7-7 p^2}} \\ \hline
AP & S_{J_zJ_z} & \displaystyle{\frac{8 (p-1) (p+1) \left(105 p^{14}-63 p^{12}-471
   p^{10}+8617 p^8-7021 p^6-3381 p^4-2597
   p^2-3381\right)}{\left(p^6+21 p^4+35 p^2+7\right)^3}} \\ \hline
LP & S_{J_zJ_z} & \displaystyle{-\frac{8 \left(24 p^8+367 p^6-1963 p^4+4921
   p^2-3381\right)}{\left(5 p^2+7\right)^3}} \\ \hline
RP & S_{J_zJ_z} & \displaystyle{\frac{8 (p-1) (p+1) \left(3 p^{10}+30 p^8-673 p^6+3321
   p^4-5614 p^2+3381\right)}{\left(p^4-2 p^2-7\right)^3}} \\ \hline \hline 
P & \langle Q \rangle & \displaystyle{4/7 } \\ \hline
AP & \langle Q \rangle & \displaystyle{\frac{4 \left(p^6+7 p^4+7 p^2+1\right)}{p^6+21 p^4+35 p^2+7} } \\ \hline
LP & \langle Q \rangle & \displaystyle{-\frac{4 \left(p^2+1\right)}{p^4-2 p^2-7} } \\ \hline
RP & \langle Q \rangle & \displaystyle{\frac{4 \left(p^2+1\right)}{5 p^2+7}} \\ \hline \hline
P & S_{QQ} &  \displaystyle{-\frac{288}{343 \left(p^2-1\right)} } \\ \hline
AP & S_{QQ} &  \displaystyle{ -\frac{32 \left(63 p^{16}+420 p^{14}+980 p^{12}+1388 p^{10}-622 p^8-1348 p^6-668 p^4-204 p^2-9\right)}{\left(p^6+21
   p^4+35 p^2+7\right)^3} } \\ \hline
LP & S_{QQ} &  \displaystyle{\frac{32 \left(2 p^{10}+7 p^8-16 p^6+58 p^4-42 p^2-9\right)}{\left(p^4-2 p^2-7\right)^3} } \\ \hline
RP & S_{QQ} &  \displaystyle{\frac{32 \left(6 p^6-7 p^4+24 p^2+9\right)}{\left(5 p^2+7\right)^3}} \\ \hline
\end{array}
\]
\caption{The analytic formulas for the case of a molecular spin $S=1$, when the state $J=3/2$ of the occupied dot lies in the bias window. 
The first column shows the type of polarizations in the leads : P for parallel ($P_L=P_R = p$), AP for anti-parallel ($P_L=-P_R=p$),
 LP ($P_L=p$, $P_R=0$) for left lead polarized only and RP ($P_R=p$, $P_L=0$) for right lead polarized only. 
The second column shows the quantity whose analytical formula is given in the third column.}
\label{table:ana1e}
\end{table*}


\begin{thebibliography}{99}
\bibitem{T-dep}
The results presented in this paper has, therefore, no $T$-dependence, and applies to the
temperature regime $\hbar\Gamma\ll k_B \Theta\ll eV$.
\end{thebibliography}
\end{document}